\documentclass[prl,showpacs,twocolumn,aps,superscriptaddress,floatfix]{revtex4}
\usepackage{amsmath}
\usepackage{amssymb}
\usepackage{bm}
\usepackage{graphicx}
\usepackage{color}
\usepackage{ulem}

\begin{document}

\title{Instabilities of the AA-stacked graphene bilayer}

\author{A.L. Rakhmanov}
\affiliation{Advanced Science Institute, RIKEN, Wako-shi, Saitama,
351-0198, Japan}
\affiliation{Institute for Theoretical and Applied Electrodynamics, Russian
Academy of Sciences, 125412 Moscow, Russia}

\author{A.V. Rozhkov}
\affiliation{Advanced Science Institute, RIKEN, Wako-shi, Saitama,
351-0198, Japan}
\affiliation{Institute for Theoretical and Applied Electrodynamics, Russian
Academy of Sciences, 125412 Moscow, Russia}

\author{A.O. Sboychakov}
\affiliation{Advanced Science Institute, RIKEN, Wako-shi, Saitama,
351-0198, Japan}
\affiliation{Institute for Theoretical and Applied Electrodynamics, Russian
Academy of Sciences, 125412 Moscow, Russia}

\author{Franco Nori}
\affiliation{Advanced Science Institute, RIKEN, Wako-shi, Saitama,
351-0198, Japan}
\affiliation{Department of Physics, University of Michigan, Ann
Arbor, MI 48109-1040, USA}

\begin{abstract}
Tight-binding calculations predict that the
AA-stacked graphene bilayer has one electron and one hole conducting bands,
and that the Fermi surfaces of these bands coincide. We demonstrate that as
a result of this degeneracy, the bilayer becomes unstable with respect to a
set of spontaneous symmetry violations. Which of the symmetries is broken
depends on the microscopic details of the system. We find that
antiferromagnetism is the more stable order parameter. This order is
stabilized by the strong on-site Coulomb repulsion. For an on-site
repulsion energy typical for graphene systems, the antiferromagnetic gap
can exist up to room temperatures.
\end{abstract}


%
%
%
%
%
%
%
%
%
%

\maketitle

\textit{Introduction}.--- Graphene is a zero-gap semiconductor demonstrating a host of unusual
electronic properties
\cite{castro_neto_review2009,chakraborty_review,meso_review}.
In recent years, the synthesis of bilayer graphene triggered investigations
of the bilayer systems, partly driven by the desire to create
graphene-based materials with an electron gap. Moreover, the graphene
bilayers are interesting materials in their own right. Most efforts have
focused on the study of the AB-stacked bilayer
\cite{mccann2006}
for which high-quality samples are available \cite{susp_bilayer2009,Mayorov_Sci2011}.
Lately, the experimental realization of the AA-stacked graphene has been reported \cite{aa_experiment2008,borysiuk_aa2011}.
In this paper we discuss electronic properties of the
AA-stacked graphene bilayer (AA-BLG), which, until recently, received very
limited theoretical attention
\cite{aa_dft2008,spin-orbit2011,borysiuk_aa2011,aa_adsorbtion2010,
aa_optics_2010}.

It is known that the AA-BLG tight-binding spectrum has four bands, of which
one electron band and one hole band cross the Fermi energy
\cite{spin-orbit2011}. The Fermi surfaces of these two bands coincide
\cite{aa_dft2008,spin-orbit2011}.
This feature has drastic consequences for the electronic properties of
the bilayer because it enables several electron and electron-phonon
instabilities, including: antiferromagnetism (AFM), current-ordered states,
bilayer exciton condensation, and instability toward the shear shift of the
layers. The type of ground state order depends on the microscopic details
of the system and can be changed by applying stress, external pressure,
the presence or absence of the substrate, etc. Below we will limit our
attention to the AFM order and the structural instability with respect to the
shear layer shift (shear instability for short). These two choices are
justified. The on-site Coulomb repulsion is the strongest interaction in
the AA-BLG system, and this interaction is sufficient to guarantee the
stability or metastability of the AFM order.
As for the shear instability, there are experimental
\cite{rotational_disorder2008,rot_disord2008_2}
and numerical
\cite{berashevich2011}
suggestions that AA-stacked graphene multilayers may be unstable with
respect to the mechanical displacement of the layers with respect to each
other. However, our calculations show that the shear instability driven by
the conducting electrons seems to have a crossover temperature which is too low to be
experimentally observable.

\textit{The model}.--- In the AA-BLG, carbon atoms of the upper layer are
located on top of the equivalent atoms of the bottom layer. The system
is modeled
by the tight-binging Hamiltonian for $p_z$ electrons of carbon atoms
\begin{eqnarray}\label{H0}
H_0&=&-t\sum_{\langle\mathbf{nm}\rangle i\sigma}
		a^{\dag}_{\mathbf{n}i\sigma}
		b^{\phantom{\dag}}_{\mathbf{m}i\sigma}
\\
\nonumber
&&- t_0\sum_{\mathbf{n}\sigma}
	a^{\dag}_{\mathbf{n}1\sigma}
	a^{\phantom{\dag}}_{\mathbf{n}2\sigma}
- t_0\sum_{\mathbf{m}\sigma}
b^{\dag}_{\mathbf{m}1\sigma}b^{\phantom{\dag}}_{\mathbf{m}2\sigma}
\nonumber\\
&&-t_g\sum_{\langle\mathbf{nm}\rangle\sigma}\left(a^{\dag}_{\mathbf{n}1\sigma}b^{\phantom{\dag}}_{\mathbf{m}2\sigma}+a^{\dag}_{\mathbf{n}2\sigma}
b^{\phantom{\dag}}_{\mathbf{m}1\sigma}\right) + \textrm{H.c.}
\nonumber
\end{eqnarray}
Here $a^{\dag}_{\mathbf{n}i\sigma}$ and $a^{\phantom{\dag}}_{\mathbf{n}i\sigma}$ ($b^{\dag}_{\mathbf{m}i\sigma}$ and $b^{\phantom{\dag}}_{\mathbf{m}i\sigma}$) are creation and annihilation operators of an electron with spin $\sigma$ in the layer
$i=1,\,2$ on the sublattice ${\cal A}$ (${\cal B}$) at site $\mathbf{n} \in
{\cal A}$ ($\mathbf{m} \in {\cal B}$).  The amplitude $t$ ($t_0$) in
Eq.~\eqref{H0} describes the in-plane (inter-plane) nearest-neighbor
hopping, while  $t_g$
corresponds to the inter-layer next-nearest neighbor hopping.
The interplane distance in bilayer graphene $c\approx3.3$\,{\AA}
\cite{borysiuk_aa2011}, and it is larger
than the in-plane carbon-carbon distance
$a\approx1.4$\,{\AA}.
Thus, the in-plane hopping integral $t$ is larger than the interplane one,
$t_0$. For calculations we will use the characteristic values
$t\approx2.57$\,eV,
$t_0\approx0.36$\,eV, $t_g\approx-0.03$\,eV~\cite{Charlier}.
We omit next next-nearest neighbor hopping between ${\cal A}$ (${\cal B}$)
sites since the corresponding term only shifts the zero-energy level.

\begin{figure}
\centering
\includegraphics[width=0.95\columnwidth]{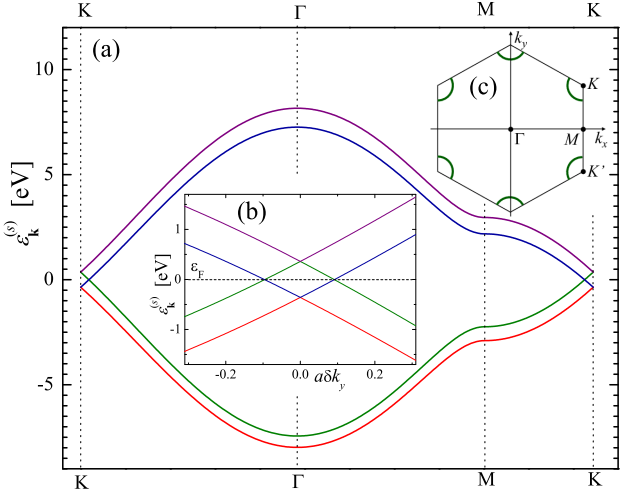}
\caption{(Color online) (a) The band structure of the AA-stacked bilayer
graphene. (b) The $\mathbf{k}$-dependence of the spectra
$\varepsilon^{(s)}_{\mathbf{k}}$ near the Dirac point
${\cal K}$
located at momentum $\mathbf{K}$;
$\mathbf{k}=\mathbf{K}+\delta k_y\mathbf{e}_y$. Bands $s=2$ and $s=3$
intersect at the Fermi level $\varepsilon_{\text{F}}$. (c) Solid (green)
lines show six
arcs of the Fermi surface in the first Brillouin zone.
\label{FigSpec0}}
\end{figure}

The elementary unit cell of bilayer graphene consists of four atoms. It
is convenient to introduce the bi-spinors
$\psi^{\dag}_{\mathbf{k}\sigma}
=
\left(
	\psi^{\dag}_{\mathbf{k}\cal{A}\sigma},
	\,\psi^{\dag}_{\mathbf{k}\cal{B}\sigma}
\right)$,
with spinor components
$\psi^{\dag}_{\mathbf{k}\cal{A}\sigma}
=
\left(
	a^{\dag}_{\mathbf{k}1\sigma},\,a^{\dag}_{\mathbf{k}2\sigma}
\right)$
and
$\psi^{\dag}_{\mathbf{k}\cal{B}\sigma}
=
e^{-i\varphi_{\mathbf{k}}}
\left(
	b^{\dag}_{\mathbf{k}1\sigma},\,b^{\dag}_{\mathbf{k}2\sigma}
\right)$,
where
$\varphi_{\mathbf{k}}=\arg \{f_{\mathbf{k}}\}$,
and
\begin{equation}\label{f}
f_{\mathbf{k}}=1+2\exp\!\!\left(\frac{3ik_xa}{2}\right)\cos\!\!\left(\frac{k_ya\sqrt{3}}{2}\right)\,.
\end{equation}
The components of the spinors
$\psi^{\dag}_{\mathbf{k}\cal{A}\sigma}$, $\psi^{\dag}_{\mathbf{k}\cal{B}\sigma}$
have different values of the sublattice index.

Let us define a set of Pauli matrices $\hat{\tau}_{\alpha}$
acting on the layer index, and a second set
$\hat{\sigma}_{\alpha}$
of Pauli matrices acting on the sublattice index. In terms of these
matrices, the bilayer Hamiltonian in
${\bf k}$-space
can be written as
$\hat{H}_{0\mathbf{k}}
=
-\left[
	t_0\hat{\tau}_{x}
	+
	(t+t_g\hat{\tau}_{x})\hat{\sigma}_{x}|f_{\mathbf{k}}|
\right]$,
or, explicitly
\begin{equation}\label{Hk}
\hat{H}_{0\mathbf{k}}=-\left(
\begin{matrix}
0&t_0&t|f_{\bf k}|&t_g|f_{\bf k}|\cr
t_0&0&t_g|f_{\bf k}|&t|f_{\bf k}|\cr
t|f_{\bf k}|&t_g|f_{\bf k}|&0&t_0\cr
t_g|f_{\bf k}|&t|f_{\bf k}|&t_0&0\cr
\end{matrix}\right).
\end{equation}
The Hamiltonian~\eqref{Hk}
is invariant under the transposition of the sublattices and of the graphene
layers. That is, $[\hat{\sigma}_x,\,\hat{H}_{0\mathbf{k}}]
=[\hat{\tau}_x,\,\hat{H}_{0\mathbf{k}}]=0$.
Thus, the eigenvectors of the matrix~(\ref{Hk})
can be classified according to the quantum numbers $\sigma$ and $\tau$,
which characterize the eigenvector parity under
$\sigma_x$ and $\tau_x$ transformations.
Using these symmetries it is easy to find the transformation which
diagonalizes
$\hat H_{0{\bf k}}$:
it is
$\hat{U}
=
\left(
	\hat{\tau}_x
	+
	\hat{\tau}_z
\right)
\left(
	\hat{\sigma}_x
	+
	\hat{\sigma}_z
\right)/2
=
\hat{U}^{-1}$.
The electron spectrum
$\varepsilon^{(s)}_{\mathbf{k}}$
 obtained consists of four bands, and each band has a unique value of the pair
$(\sigma, \tau)$:
\begin{eqnarray}
\label{E0k}
&&\varepsilon^{(1)}_{\mathbf{k}}= -t_0-(t+t_g)|f_{\bf k}|,
\quad\sigma=1, \quad\tau=1,\\
\label{band2}
&&\varepsilon^{(2)}_{\mathbf{k}}=+t_0-(t-t_g)|f_{\bf k}|,
\quad\sigma=1, \quad\tau=-1,\\
\label{band3}
&&\varepsilon^{(3)}_{\mathbf{k}}=-t_0+(t+t_g)|f_{\bf k}|,
\quad\sigma=-1, \quad\tau=1,\\
&&\varepsilon^{(4)}_{\mathbf{k}}=+t_0+(t-t_g)|f_{\bf k}|,
\quad\sigma=-1, \quad\tau=-1.
\end{eqnarray}
The band structure is shown in
Fig.~\ref{FigSpec0}.
The bands $s=2$ and
$s=3$
cross the Fermi energy level near the Dirac point
${\cal K}$,
located at momentum
$\mathbf{K}=2\pi\{\sqrt{3},\,1\}/(3\sqrt{3}a)$
and the Dirac point
${\cal K}'$
located at momentum
$\mathbf{K}'=2\pi\{\sqrt{3},\,-1\}/(3\sqrt{3}a)$
[see
Fig.~\ref{FigSpec0}(b)].
The most interesting feature of this band structure is that at half filling
(which corresponds to undoped AA-BLG) the Fermi surfaces of both bands
coincide. The
Fermi level is $\varepsilon_{\text{F}}=t_gt_0/t$,
while the Fermi surfaces are given by the equation
$|f_{\bf k}|=t_0/t$. For
$t_0/t \ll 1$
one can expand the function
$|f_{\bf k}|$
near the Dirac points and demonstrate that the Fermi surface
consists of six arcs inside the first Brillouin zone with the radius
$k_{r}=2t_0/(3ta)$ [Fig.~\ref{FigSpec0}(c)].

The matching of the Fermi surfaces turns out to be quite stable against
changes in the tight-binding Hamiltonian. First, it survives if
we add more distant hopping terms to
$H_0$. Moreover, even layer-asymmetric systems (e.g., similar to the
single-side hydrogenated graphene
\cite{sshg})
may posses this property. However, it is clear that the different
types of interactions, e.g., electron-electron or electron-phonon ones, can
destabilize such a degenerate spectrum.

\textit{Mean-field Hamiltonian}.--- The presence of two bands with identical Fermi surfaces makes the system
unstable with respect to spontaneous symmetry
breaking. We will demonstrate that the Hamiltonian symmetries
$\sigma_x$ and $\tau_x$ can be used to narrow the possible symmetry choices.

In the mean-field approach, the two-particle interaction operator
$H_{\rm int}\propto\psi^\dag\psi^\dag\psi\psi$
is replaced by a single-particle operator
$\delta H_{\rm int}\propto\langle\psi^\dag\psi\rangle\psi^\dag\psi$,
where the average $\langle\psi^\dag\psi\rangle$ is different
types of non-superconducting order parameter. The values of these order
parameters are found from the self-consistency conditions.
To be at least metastable, the order parameter must open a gap at the Fermi
level. The most general form of $\delta H_{\rm int}$, which can open an
insulating gap, is
\begin{eqnarray}
\delta H_{\text{int}}
&=&
\sum_{\mathbf{k}\sigma}
	\psi^{\dag}_{\mathbf{k}\sigma}
	\delta\hat{H}_{\mathbf{k}\sigma}
	\psi^{\phantom{\dag}}_{\mathbf{k}\sigma}\,,
\label{Delta}
\\
\delta\hat{H}_{\mathbf{k}\sigma}
&=&
\sum_{\alpha}
	\left(
		\Delta^{\alpha}_{AB\mathbf{k}\sigma}
		\hat{\sigma}_{\alpha}
		+
		\Delta^{\alpha}_{12\mathbf{k}\sigma}
		\hat{\tau}_{\alpha}
	\right)%
+
\sum_{\alpha\beta}
	\Delta^{\alpha\beta}_{\mathbf{k}\sigma}
	\hat{\tau}_{\alpha}
	\hat{\sigma}_{\beta}
\,,
\nonumber
\end{eqnarray}
where $\Delta^{\alpha}_{AB\mathbf{k}\sigma}$,
$\Delta^{\alpha}_{12\mathbf{k}\sigma}$, and
$\Delta^{\alpha\beta}_{\mathbf{k}\sigma}$ are real-valued order parameters,
which, in general, are functions of $\mathbf{k}$.
To open a gap, the corresponding term in
$\delta H_{\rm int}$ must couple the conducting bands
$\varepsilon^{(2)}_{\mathbf{k}}$
and $\varepsilon^{(3)}_{\mathbf{k}}$.
Since these bands have unequal values of $\sigma$ and $\tau$ [see
Eq.~(\ref{band2}) and Eq.~(\ref{band3})],
therefore, only terms containing $\Delta^{\alpha\beta}_{\mathbf{k}\sigma}$
with $\alpha,\beta\neq x$ may couple these bands.
Other terms commute either with $\sigma_x$ or $\tau_x$.

To find the renormalized spectrum of the bands near the Fermi-level, we
should diagonalize the matrix
$\hat{H}_{\mathbf{k}\sigma}=\hat{H}_{0\mathbf{k}}+\delta\hat{H}_{\mathbf{k}\sigma}$.
Performing the unitary transformation of $\hat{H}_{\mathbf{k}\sigma}$ with
$\hat{U}$ written above, we obtain \begin{equation}\label{HDelta}
\hat{U}^{-1}\hat{H}_{\mathbf{k}\sigma}\hat{U}=\left(
\begin{matrix}
\ddots&\cdots&\cdots&\cdots\\
\vdots&\varepsilon^{(2)}_{\mathbf{k}}+\delta\varepsilon^{(2)}_{\mathbf{k}\sigma}&\Delta_{\mathbf{k}\sigma}&\vdots\\
\vdots&\Delta^{*}_{\mathbf{k}\sigma}&\varepsilon^{(3)}_{\mathbf{k}}+\delta\varepsilon^{(3)}_{\mathbf{k}\sigma}&\vdots\\
\cdots&\cdots&\cdots&\ddots
\end{matrix}\right),
\end{equation}
where
$\delta\varepsilon^{(2)}_{\mathbf{k}\sigma}=\Delta^{x}_{AB\mathbf{k}\sigma}-\Delta^{x}_{12\mathbf{k}\sigma}-\Delta^{xx}_{\mathbf{k}\sigma}$,
$\delta\varepsilon^{(3)}_{\mathbf{k}\sigma}=-\Delta^{x}_{AB\mathbf{k}\sigma}+\Delta^{x}_{12\mathbf{k}\sigma}-\Delta^{xx}_{\mathbf{k}\sigma}$,
and
$\Delta_{\mathbf{k}\sigma}=\Delta^{zz}_{\mathbf{k}\sigma}+\Delta^{yy}_{\mathbf{k}\sigma}+i(\Delta^{zy}_{\mathbf{k}\sigma}+\Delta^{yz}_{\mathbf{k}\sigma})$.
Other elements of this matrix are unimportant for further consideration.
Solving the secular equation for the $2\times2$ matrix presented in
Eq.~\eqref{HDelta}, we obtain the renormalized spectrum of the bands with
$s=2,3$:
\begin{eqnarray}\label{SpecDelta}
E^{(2,3)}_{\mathbf{k}\sigma}&=&\frac{1}{2}\left(\varepsilon^{(2)}_{\mathbf{k}}+\delta\varepsilon^{(2)}_{\mathbf{k}\sigma}+%
\varepsilon^{(3)}_{\mathbf{k}}+\delta\varepsilon^{(3)}_{\mathbf{k}\sigma}\right)
\mp\\%
&&\frac{1}{2}\sqrt{\left(\varepsilon^{(2)}_{\mathbf{k}}+\delta\varepsilon^{(2)}_{\mathbf{k}\sigma}-%
\varepsilon^{(3)}_{\mathbf{k}}-\delta\varepsilon^{(3)}_{\mathbf{k}\sigma}\right)^2+4\left|\Delta_{\mathbf{k}\sigma}\right|^2}\,.\nonumber
\end{eqnarray}
The gap between renormalized bands is equal to
$\Delta_{0}=2\min_{\mathbf{k}}\left|\Delta_{\mathbf{k}\sigma}\right|$. We
see that the contribution to the gap comes only from
$\Delta^{zz}_{\mathbf{k}\sigma}$, $\Delta^{yy}_{\mathbf{k}\sigma}$,
$\Delta^{zy}_{\mathbf{k}\sigma}$, and $\Delta^{yz}_{\mathbf{k}\sigma}$
order parameters which break down both sublattice and layer symmetries.  It
is easy to show that other elements of the full $4\times4$
matrix~\eqref{HDelta} give only the second-order contribution to this
result.  Taking other order parameters zero, the matrix
$\delta\hat{H}_{\mathbf{k}\sigma}$ can be written in the form
\begin{equation}\label{Hgap}
\delta\hat{H}_{\mathbf{k}\sigma}=\left(
\begin{matrix}
\Delta^{zz}_{\mathbf{k}\sigma}&-i\Delta^{yz}_{\mathbf{k}\sigma}&-i\Delta^{zy}_{\mathbf{k}\sigma}&-\Delta^{yy}_{\mathbf{k}\sigma}\\
i\Delta^{yz}_{\mathbf{k}\sigma}&-\Delta^{zz}_{\mathbf{k}\sigma}&\Delta^{yy}_{\mathbf{k}\sigma}&i\Delta^{zy}_{\mathbf{k}\sigma}\\
i\Delta^{zy}_{\mathbf{k}\sigma}&\Delta^{yy}_{\mathbf{k}\sigma}&-\Delta^{zz}_{\mathbf{k}\sigma}&i\Delta^{yz}_{\mathbf{k}\sigma}\\
-\Delta^{yy}_{\mathbf{k}\sigma}&-i\Delta^{zy}_{\mathbf{k}\sigma}&-i\Delta^{yz}_{\mathbf{k}\sigma}&\Delta^{zz}_{\mathbf{k}\sigma}
\end{matrix}\right).
\end{equation}

In principle there is a huge number of possible instabilities generated by different types of interactions which can lead to the mean-field interaction Hamiltonian of the form in Eq.~\eqref{Hgap}. Below we will show that $\Delta^{zz}_{\mathbf{k}\sigma}$ can be related to the G-type antiferromagnetic order parameter (i.e., each spin is antiparallel to all nearest-neighboring spins) produced by the on-site Coulomb repulsion. The $\Delta^{yy}_{\mathbf{k}\sigma}$ can be attributed to the instability toward the homogeneous shift of the one graphene layer with respect to another one. The order parameters $\Delta^{zy}_{\mathbf{k}\sigma}$ and $\Delta^{yz}_{\mathbf{k}\sigma}$ can correspond, e.g., to excitons, which produce a current flowing inside and between the layers, respectively.

\textit{Antiferromagnetic state}.--- It is known that the Coulomb
interaction among electrons in graphene is rather strong and the value of
the on-site Coulomb repulsion energy $U$ is about $10$\,eV~\cite{Wehling}.
However, graphene remains semimetal since the electron density of states at
the Fermi level is zero. In contrast, the AA-BLG has a Fermi surface and the
density of states at the Fermi level is finite. Then, one can expect that
the role of electron-electron interactions in AA-BLG is more important and
it can affect the ground state. We restrict ourselves here by considering
the on-site Coulomb interaction and write the Hubbard Hamiltonian in the form
\begin{equation}\label{U}
H_{\text{int}}=
\frac{U}{2}\sum_{\mathbf{n}i\sigma}
n_{\mathbf{n}i\cal{A}\sigma}n_{\mathbf{n}i\cal{A}\bar{\sigma}}+
\frac{U}{2}\sum_{\mathbf{m}i\sigma}
n_{\mathbf{m}i\cal{B}\sigma}n_{\mathbf{m}i\cal{B}\bar{\sigma}}\,,
\end{equation}
where $n_{\mathbf{n}i\cal{A}\sigma}=
a^{\dag}_{\mathbf{n}i\sigma}a^{\phantom{\dag}}_{\mathbf{n}i\sigma}$,
$n_{\mathbf{m}i\cal{B}\sigma}=b^{\dag}_{\mathbf{m}i\sigma}
b^{\phantom{\dag}}_{\mathbf{m}i\sigma}$, and $\bar{\sigma}=-\sigma$. It is
known that the ground state of the Hubbard-like models at half-filling can
be antiferromagnetic (AFM). For the AA-BLG symmetry, three types of AFM
ordering (having different spin arrangement inside the unit cell) are
possible. However, only the G-type AFM order (AFM arrangement both between
sublattices and layers) opens a gap at the Fermi level even if the interaction
is arbitrary small.

In the mean-field approximation we represent
$n_{\mathbf{n}ia\sigma}$ ($a={\cal A},\,{\cal B}$) in Eq.~\eqref{U} in the
form $n_{\mathbf{n}ia\sigma}=n_{ia\sigma}+\delta n_{\mathbf{n}ia\sigma}$,
where $n_{ia\sigma}=\langle n_{\mathbf{n}ia\sigma}\rangle$ and $\delta
n_{\mathbf{n}ia\sigma}=n_{\mathbf{n}ia\sigma}-n_{ia\sigma}$. The mean-field
Hamiltonian is obtained then by neglecting the terms quadratic in $\delta
n_{\mathbf{n}ia\sigma}$.
For G-type AFM, the spin-up and spin-down electron densities are
redistributed as
$n_{1\cal{A}\uparrow}=n_{2\cal{B}\uparrow}=n_{2\cal{A}\downarrow}=n_{1\cal{B}\downarrow}=(1+\Delta
n)/2$ and
$n_{1\cal{A}\downarrow}=n_{2\cal{B}\downarrow}=n_{2\cal{A}\uparrow}=n_{1\cal{B}\uparrow}=(1-\Delta
n)/2$, while the total on-site electron density
$n_{ia\sigma }
+
n_{ia\bar\sigma }$
remains equal to unity. Thus,
the mean-field interaction Hamiltonian has the form in Eq.~\eqref{Delta} with
$\Delta^{zz}_{\mathbf{k}\uparrow}=-\Delta$,
$\Delta^{zz}_{\mathbf{k}\downarrow}=+\Delta$, where 
$\Delta=U\Delta n/2$.
Other terms in Eq.~\eqref{Delta} are equal to zero.

\begin{figure}
\centering
\includegraphics[width=0.95\columnwidth]{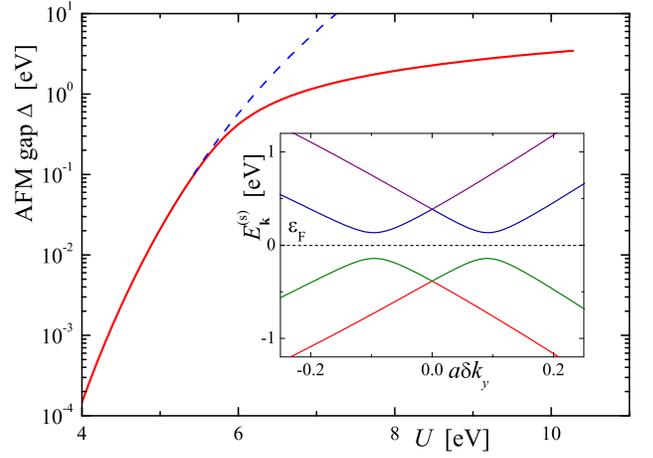}
\caption{(Color online) The dependence of the AFM gap $\Delta$ on the
on-site Coulomb repulsion $U$. Solid (red) curve is calculated by solving
Eq.~\eqref{DeltaEq2}, while the dashed (blue) curve is calculated from
Eq.~\eqref{DeltaAFM}. The inset shows the electron spectrum near
$\mathcal{K}$ point at $U=5.5$\,eV
($\Delta\approx0.12$\,eV).\label{FigDelta}} \end{figure}

The eigenvalues $E^{(s)}_{\mathbf{k}\sigma}$ and eigenvectors
$\upsilon^{(s)}_{p\mathbf{k}\sigma}$ of the matrices
$\hat{H}_{\mathbf{k}\sigma}=\hat{H}_{0\mathbf{k}}+\delta\hat{H}_{\mathbf{k}\sigma}$
can be found analytically. The spectra of spin-up and spin-down electrons
are equal. The spectra of the bands $2$ and $3$ have the form in
Eq.~\eqref{SpecDelta} with
$\delta\varepsilon^{(2)}_{\mathbf{k}\sigma}=\delta\varepsilon^{(3)}_{\mathbf{k}\sigma}=0$,
and $\left|\Delta_{\mathbf{k}\sigma}\right|=\Delta$. The spectra of the
bands $1$ and $4$ are given by Eq.~\eqref{SpecDelta} with the replacement
$\varepsilon^{(2,3)}_{\mathbf{k}}\to\varepsilon^{(1,4)}_{\mathbf{k}}$.
When the gap is open, the lower two bands are filled, while the upper two
are empty. To find the value of the gap, one needs to solve the
self-consistent equation for $\Delta n=2\Delta/U$:
\begin{eqnarray}\label{DeltaEq1}
&&n_{1\cal{A}\uparrow}=\frac12+\frac{\Delta}{U}=%
\sum_{s=1,2}\int\!\!\frac{d\mathbf{k}}{V_{\text{BZ}}}\left|\upsilon^{(s)}_{1\mathbf{k}\uparrow}\right|^2=\frac12+\\%
&&\frac{1}{4}\!\!\int\!\!\frac{d\mathbf{k}}{V_{\text{BZ}}}\!\!\left[\frac{\Delta}{\sqrt{\Delta^2+(t|f_{\mathbf{k}}|+t_0)^2}}+%
\frac{\Delta}{\sqrt{\Delta^2+(t|f_{\mathbf{k}}|-t_0)^2}}\right],\nonumber
\end{eqnarray}
where $V_{\text{BZ}}=8\pi^2/(3\sqrt{3}a^2)$ is the volume of the first
Brillouin zone. We introduce the dimensionless
density of states $\rho_0(\zeta)=\int
d\mathbf{k}\,\delta(\zeta-|f_{\mathbf{k}}|)/V_{\text{BZ}}$, which is
related~\cite{castro_neto_review2009} to the graphene density of states
$\rho_{\text{gr}}(E)$ according to $\rho_{\text{gr}}(E)=\rho_{0}(|E/t|)/t$.
Equation~\eqref{DeltaEq1} then becomes
\begin{equation}\label{DeltaEq2}
\int\limits_0^3\!\!d\zeta\!\!\left[\frac{\rho_0(\zeta)}{\sqrt{\delta^2+(\zeta+\zeta_0)^2}}+%
\frac{\rho_0(\zeta)}{\sqrt{\delta^2+(\zeta-\zeta_0)^2}}\right]\!\!=\frac{4t}{U}\,,
\end{equation}
where $\delta=\Delta/t$ and $\zeta_0=t_0/t$. The integral of the second
term in the left-hand side of Eq.~\eqref{DeltaEq2} diverges logarithmically
when $\Delta\to0$. In the limit of small $\Delta$, from
Eq.~\eqref{DeltaEq2} one can derive
\begin{equation}\label{DeltaAFM}
\Delta=2\sqrt{t_0(3t-t_0)}\exp\left\{-\frac{4t-U\eta(\zeta_0)}{2U\rho_0(\zeta_0)}\right\}\,,
\end{equation}
where
\begin{equation}
\eta(\zeta_0)=\int\limits_0^3\!\!d
\zeta\left[		\frac{\rho_0(\zeta)}{\zeta+\zeta_0}
		+		\frac{\rho_0(\zeta)-\rho_0(\zeta_0)}
		     {\left|\zeta-\zeta_0\right|}	\right].
\end{equation}

Figure~\ref{FigDelta} shows the dependence of $\Delta$ on $U$.
Taking the value of $U=8\div 9$\,eV~\cite{Wehling}, we obtain
$\Delta\cong2\div3$\,eV and the magnetic moment at each site $\mu_{\text{B}}\Delta n$ about 1$\mu_{\text{B}}$.
However, we do not know exact value of $U$ for AA-BLG.
These calculations were done at zero temperature. At finite temperatures no long-range AFM
order exists. The crossover temperature $T^{*}$ between the short-range AFM
state and paramagnetic state can be estimated as
$T^{*}\sim\Delta/k_{\text{B}}$.

\textit{Discussion}.--- Other possible types of ordering could be
considered following the same approach used for AFM.  However, whether a
particular order is stable and observable depends on the values of the hopping
amplitudes and a characteristic energy of the appropriate interaction.
For example, applying the mean field approximation to the model with the
on-site repulsion we find that the charge density is unstable for our
choice of parameters.

The next evident possibility to open a gap in the spectrum is to shear one
graphene layer with respect to another.  The shift $u$ deforms the shape of
the unit cell changing ${\cal A}{\cal B}$ bonds between different layers,
giving rise to the appearance of the order parameter
$\Delta_{\mathbf{k}\sigma}^{yy}$. Assuming that the hopping amplitude $t_g$
changes linearly with $u$, we can write for different bonds $t_g(u)\approx t_g(0)\pm(\partial t_g/\partial u)u$. Now
the electronic energy of the system becomes a function of $u$. Taking into
account the elastic contribution $C_{\textrm{sh}}u^2/2$ (where $C_{\textrm{sh}}$ is the
corresponding shear modulus) to the total energy and minimizing
this energy with respect to $u$, we obtain the value of the equilibrium shift between the layers
\begin{equation}\label{shear}
u_{\textrm{eq}}\approx \frac{t}{|\partial t_g/\partial u|}\exp\left[-\frac{\pi C_{\textrm{sh}}a^2t^2}{(\partial t_g/\partial u)^2ct_0}\right].
\end{equation}
If we assume that $C_{\textrm{sh}}$ is approximately equal to the shear modulus in graphite and
$|\partial t_g/\partial u|\sim t_g/a$,
we conclude that the shift $u_{\textrm{eq}}$ and the corresponding energy gain are too small
to be observable. However, this conclusion must be taken cautiously. First,
we have no accurate information on $C_{\textrm{sh}}$ and
$|\partial t_g/\partial u|$, whose precise values are very important for
the estimate of
$u_{\textrm{eq}}$
and the corresponding energy. Moreover,
$u_{\textrm{eq}}$
could be enhanced by pressure or the presence of a substrate. Finally, the
shift can be induced by a different mechanism.

In conclusion, we demonstrate that the AA-BLG is unstable with respect to a
set of symmetry-breaking instabilities, which can give rise to the
existence of several order parameters of different nature. We show that the AFM
order can be observed in the system. The possible existence of other types of orders in the AA-BLG
depends on the system parameters and the external conditions (temperature,
pressure, substrate, etc).

We thank L.~Openov for stimulating discussions.
This work was supported in part by JSPS-RFBR Grant
No.~09-02-92114, RFBR Grant No.~09-02-00248, LPS, NSA, ARO, NSF grant No.~0726909,
Grant-in-Aid for Scientific Research~(S), MEXT Kakenhi on Quantum
Cybernetics, and the JSPS via its FIRST program.
AOS acknowledges partial support from the Dynasty Foundation.

\vspace*{-0.1in}
%


%

\end{document}